\documentclass[twocolumn,showpacs,superscriptaddress,prl]{revtex4-1}
\usepackage{graphicx}
\usepackage{color}
\usepackage{amsmath,amsthm,amssymb}
\usepackage{braket}
\usepackage[colorlinks,citecolor=blue,linkcolor=red]{hyperref}

\begin{document}
 
\title{Competing Incommensurate Spin Fluctuations and Magnetic Excitations in Infinite-Layer Nickelate Superconductors}

\author{Christopher Lane}
\email{laneca@lanl.gov}
\affiliation{Theoretical Division, Los Alamos National Laboratory, Los Alamos, New Mexico 87545, USA}

\author{Ruiqi Zhang}
\affiliation{Department of Physics \& Engineering Physics, Tulane University, New Orleans, Louisiana 70118, USA}

\author{Bernardo Barbiellini}
\affiliation{School of Engineering Science, 
Lappeenranta-Lahti University of Technology (LUT), Lappeenranta, Finland}
\affiliation{Department of Physics, Northeastern University, Boston, MA 02115, USA}

\author{Robert S. Markiewicz}
\affiliation{Department of Physics, Northeastern University, Boston, MA 02115, USA}

\author{Arun Bansil}
\affiliation{Department of Physics, Northeastern University, Boston, MA 02115, USA}

\author{Jianwei Sun}
\affiliation{Department of Physics \& Engineering Physics, Tulane University, New Orleans, Louisiana 70118, USA}

\author{Jian-Xin Zhu}
\affiliation{Theoretical Division, Los Alamos National Laboratory, Los Alamos, New Mexico 87545, USA}

\date{\today} 
\begin{abstract}
The recently discovered infinite-layer nickelates show great promise in helping to disentangle the various cooperative mechanisms responsible for high-temperature superconductivity. However, lack of antiferromagnetic order in the pristine nickelates presents a challenge for connecting the physics of the cuprates and nickelates. Here, by using a quantum many-body Green's function-based approach to treat the electronic and magnetic structures, we unveil the presence of many two- and three-dimensional magnetic stripe instabilities that are shown to persist across the phase diagram of LaNiO$_2$. Our analysis indicates that the magnetic properties of the infinite-layer nickelates are closer to those of the doped cuprates which host inhomogeneous ground states rather than the undoped cuprates. The computed magnon spectrum in LaNiO$_2$ is found to contain an admixture of contributions from localized and itinerant carriers. The theoretically obtained magnon dispersion is in accord with the results of the corresponding RIXS experiments. Our study gives insight into the origin of inhomogeneity in the infinite-layer nickelates and their relationship with the cuprates.
\end{abstract}

\pacs{}

\maketitle 

\section{Introduction}
The common thread linking the family of high-temperature superconductors is that competing interactions involving charge, spin, lattice and orbital degrees of freedom conspire with electronic correlation effects to produce many complex properties in this materials family. \cite{dagotto2005complexity}. The phase diagrams of transition-metal oxides, for example, are astonishingly complex and exhibit unconventional superconducting pairing, pseudogap and glassy phases, and colossal magnetoresistance in stark contrast to the standard metals\cite{keimer2015quantum,dagotto2003nanoscale,platzman1973waves,sobota2021angle}. The high-T$_c$ cuprates have been of intense interest where a wide variety of experiments have reported the presence of competing and intertwined inhomogeneous orders that could contribute to the pairing mechanism \cite{fradkin2015colloquium,he2018rapid,li2018coherent,jiang2022stripe}. However, deconstructing the mechanism of high-Tc superconductivity and possible contributions involved in the pairing process has remained a challenge. Understanding electronic and magnetic properties of materials related to the cuprates can provide insight into the mechanism of superconductivity in the cuprates and other high-Tc superconductors. 

Recently, superconductivity was discovered in doped infinite-layer nickelates\cite{li2019superconductivity}. The $R$NiO$_2$ ({\it R}$=$Nd,Pr,La) family of compounds is isostructural to the infinite-layer parent cuprate CaCuO$_2$ \cite{botana2020similarities}, where the 2D NiO$_2$ planes are separated by rare earth spacer layers\cite{osada2020superconducting,osada2021nickelate,zeng2020phase}. Due to the missing apical oxygen in $R$NiO$_2$, the nickel atoms take a $3d^9$ configuration that is equivalent to Cu$^{2+}$ in the cuprates, thereby strengthening the hypothesis that these two materials families are electronically analogous\cite{anisimov1999electronic,lee2004infinite}.

To date, a number of experimental techniques have been employed to elucidate possible connections between the nickel and copper based superconductivity, and the role of competing orders in their phase diagrams. Transport measurements on the nickelates report key departures from the cuprate phenomenology. Specifically, no Mott or antiferromagnetic (AFM) parent phase is observed, with both the underdoped and overdoped nickelates only displaying a weakly insulating state\cite{li2020superconducting,osada2020phase,zeng2020phase,osada2021nickelate,zeng2022superconductivity,lee2022character}. The Hall coefficient $R_{H}$ is found to change sign at optimal doping ($x\sim0.17$) in all nickelates that have been investigated, signaling the existence and importance of both hole and electron pockets at the Fermi level\cite{li2020superconducting,osada2020phase,zeng2020phase,osada2021nickelate,zeng2022superconductivity,lee2022character}. X-ray absorption spectroscopy (XAS) and resonant inelastic X-ray scattering (RIXS) experiments find the doped holes to reside on the Ni-$d_{x^2-y^2}$ orbtials, with minor $5d$ electron doping due to Ni-$3d$/La-$5d$ hybridization\cite{hepting2020electronic,rossi2021orbital}, suggesting the coexistence of multiple active orbitals at the Fermi level. Despite these differences, the nickelates exhibit signatures of stripe formation and singlets in the ground state, similar to the underdoped curpates\cite{rossi2021orbital,zeng2020phase,rossi2021broken,tam2021charge}.

\begin{figure*}[t!]
\includegraphics[width=0.99\textwidth]{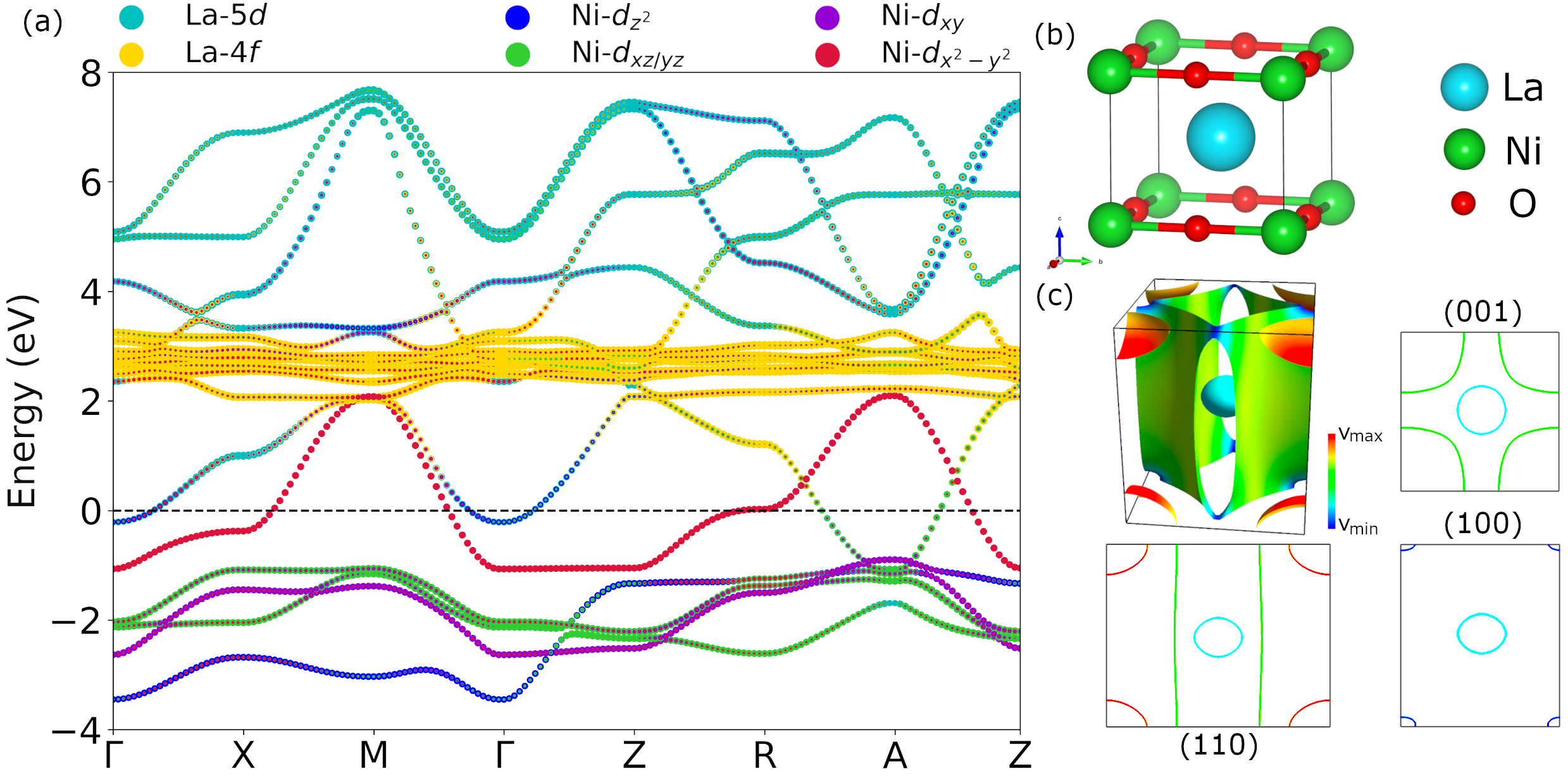}
\caption{(color online) (a) Electronic band dispersion in LaNiO$_2$ in the NM phase. Size and color of the dots is proportional to the fractional weight of the various indicated site-resolved orbital projections. (b) Primitive crystal structure of NM LaNiO$_2$. (c) Calculated Fermi surface of LaNiO$_2$ and its cuts in the (001), (100), and (110) planes. Colors on various Fermi surface sheets indicate the magnitude of the associated Fermi velocities, as specified by the colorbar. } 
\label{fig:bands}
\end{figure*}

Magnetic properties of the nickelates differ substantially from those of the cuprates in that the pristine undoped cuprates exhibit commensurate AFM order. In sharp contrast, no long range AFM order is found in $R$NiO$_2$ \cite{hayward1999sodium,hayward2003synthesis,ikeda2016direct,crespin1983reduced}, despite the existence of robust two-dimensional (2D) AFM spin wave dispersions observed in RIXS\cite{lu2021magnetic}. Instead, strong non-local magnetic correlations and weak to intermediate glassy short range behavior appears to dominate the ground state\cite{ortiz2022magnetic,lin2022universal,ikeda2016direct}. This suggests that the magnetic properties of the infinite-layer nickelates are closer to those of the doped cuprates where inhomogeneities comprise the ground state, rather than the undoped cuprates which present a pristine ordered phase.  

As a coherent picture of the magnetism in infinite-layer nickelates has started to emerge, multiple DFT\cite{botana2020similarities,wan2021exchange,liu2020electronic,jung2020antiferromagnetic,zhang2021magnetic} and DFT+DMFT\cite{ortiz2022magnetic,wan2021exchange,lechermann2021doping,gu2020substantial} studies have obtained a magnetic ground state, which is at odds with the experimental evidence, which consistently finds a landscape of nearly degenerate spin orders, indicating that magnetic correlations in infinite-layer nickelates are frustrated, giving way to numerous competing magnetic configurations. In this connection, a variety of tight-binding models have been invoked~\cite{Sakakibara2019,Gu2020,Kitatani2020,zhou2020spin,lechermann2020multiorbital} to understand the low-energy physics and magnetic instabilities in the nickelates. These models, however, are fundamentally limited because the relevant orbitals involved remain unclear.  

In this article, we propose that the absence of long-range order in the infinite-layer nickelates originates from the a large number of competing symmetry breaking magnetic instabilities in the pristine phase, which strongly couple at low temperatures as is common in glassy phases of matter. A 3D-2D crossover in the magnetic fluctuation spectrum takes place near start of the  superconducting dome in doping. By using the C-type AFM state as a model system, we obtain the spin wave spectrum, where the spin wave dispersion is in accord with the corresponding results of RIXS experiments. An analysis of the spin wave spectrum reveals a dominant Ni-$d_{x^2-y^2}$ contribution and an admixture of contributions from Ni-$d_{xy}$ and La-$4f$ orbitals, indicating the need to go beyond the commonly used one band paradigm for accurate modeling of magnetic excitations in the nickelates.

\section{Computational Details}
{\it Ab initio} calculations were carried out by using the pseudopotential projector-augmented wave method\cite{Kresse1999} implemented in the Vienna {\it ab initio} simulation package (VASP) \cite{Kresse1996,Kresse1993} with an energy cutoff of $520$ eV for the plane-wave basis set. Exchange-correlation effects were treated by using the strongly constrained and appropriately normed (SCAN) meta-GGA scheme\cite{Sun2015}. A 15 $\times$ 15 $\times$ 17  $\Gamma$-centered $k$-point mesh was used to sample the Brillouin zone. Spin-orbit coupling effects were included self-consistently. We used the experimental low-temperature P4/mmm crystal structure to
initialize our computations\cite{hayward2003synthesis}. All atomic sites in the NM and C-AFM unit cells along with the cell dimensions were relaxed using a conjugate gradient algorithm to minimize the energy with an atomic force tolerance of $0.01$ eV/\AA .  A total energy tolerance of $10^{-5}$ eV was used to determine the self-consistent charge density. The many-body theory calculations of the spin-orbital fluctuations and magnetic excitations were performed by employing a real-space tight-binding model Hamiltonian, which was obtained by using the VASP2WANNIER90~\cite{Pizzi2020} interface. For LaNiO$_2$, the full manifold of Ni-3$d$, La-5$d$, and La-4$f$ states was included in generating the Wannier functions, whereas in the C-AFM phase only $d_{z^2}$, $d_{xy}$, $f_{z^3}$, and $f_{x(x^2-3y^2)}$ were retained on the La sites. In the NM state, the response functions were evaluated over a $51\times 51\times 51$ $k$-mesh at $0.001$ K, where only the 4 bands at the Fermi level were used in the summation over the bands in the Lindhard susceptibility. In the C-AFM phase, a $23\times 23\times 23$ $k$-mesh was employed, where all the bands were used in the sum over eigenvalues along with a $\omega$- and $X$-mesh used for the binning method outlined in the Appendix that span $[0,1]$ eV and $[-10,10]$ eV of density $N_\omega=101$ and $N_X=2001$, respectively, along with a finite broadening of $0.01$ eV. Coulomb interaction was included following Ref.~\onlinecite{lane2022identifying}, where we used an on-site Coulomb potential of $1.22$ eV and $3.76$ eV to access the leading Stoner instability in the NM phase and the magnon dispersion in the AFM state, respectively.

\section{Electronic Structure and Competing Fluctuations}
Figure~\ref{fig:bands} (a) shows the Wannier-interpolated electronic band structure for LaNiO$_2$ in the NM phase [Fig.~\ref{fig:bands} (b)]. In the lanthanum-based infinite-layer nickelates three distinct bands are found to cross the Fermi level: one of nearly pure Ni-$3d_{x^2-y^2}$ character, while the other two are derived from Ni-$3d_{xy/yz}$ and Ni-$3d_{z^2}$/La-$5d$ orbitals, respectively. The latter bands give rise to spherical electron Fermi pockets at $\Gamma$ and A symmetry points in the Brillouin zone [Figure~\ref{fig:bands} (c)], whereas the former band generates a large, slightly warped quasi-2D cylindrical Fermi surface similar to the cuprates. The relatively isolated Ni-$3d_{x^2-y^2}$ state is the result of the oxygen deintercalation  process used to convert $R$NiO$_3$ to $R$NiO$_2$,  thereby reorganizing the electronic states from that of an octahedral crystal field to a square-planar geometry. Notably, the half-filled Ni-${d}_{x^2-y^2}$ band closely resembles the corresponding band in the cuprates\cite{lane2018antiferromagnetic}, except for an enhanced $k_z$ dispersion due to the three-dimensional (3D) nature of the crystal structure. This results in a shift in the position of the van Hove singularity (VHS) from below to infinitesimally above the Fermi level along the $k_z$ direction in the Brillouin zone.  Concomitantly, the Fermi surface transitions from being hole-like (open) in the $k_z=0$ plane to becoming electron-like (closed) in the $k_z=\pi/c$ plane.

\begin{figure*}[t!]
\includegraphics[width=0.99\textwidth]{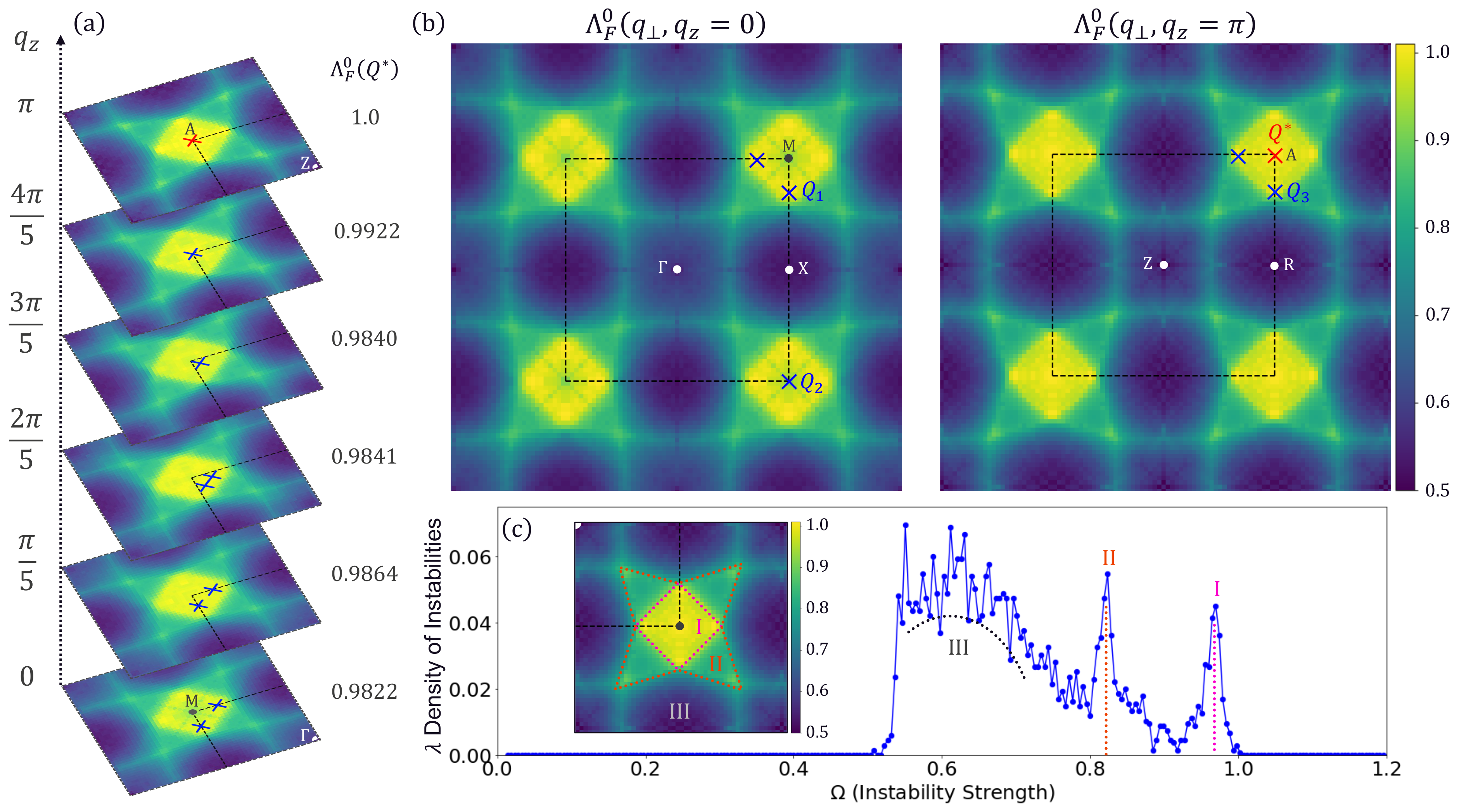}
\caption{(color online) Momentum dependence of the largest instability $\Lambda^{0}_{F}(\mathbf{q},\omega=0)$ for pristine LaNiO$_2$ in the NM phase for various slices along $q_z$ (a) and in the $\Gamma$ and $Z$ planes (b) of the Brillouin zone. The red and blue `x' mark the maximal and sub-maximal instability momenta $\mathbf{Q}^{*}$, respectively. The black dashed line denotes the boundary of the Brillouin zone. (c) The density of instabilities with the two Van Hove-like singularities and step edge indicated by Roman numerals $\mathbb{I}$, $\mathbb{II}$, and $\mathbb{III}$. The inset in (c) shows the regions of origin in $\Lambda^{0}_{F}(\mathbf{q},\omega=0)$ of the various marked peaks.} 
\label{fig:flucutations}
\end{figure*}

A key difference between the parent compounds of the cuprates and nickelates is the lack of long-range magnetic order in the nickelates. Instead, strong AFM correlations and glassy dynamics are observed to dominate throughout the phase diagram of $R$NiO$_2$. To gain insight into the landscape of charge and magnetic instabilities in the ground state, we examine the response $(\delta\rho)$ of the system to an infinitesimal perturbing source field $(\delta \pi)$. The associated response function is given by,
\begin{align}\label{eq:general_response}
\chi^{IJ}(1,2)&=\frac{\delta\rho^I(1)}{\delta \pi^{J}(2)}\\
&=\chi^{IJ}_0(1,2)+\chi^{IM}_0(1,3)v^{ML}(3,4)\chi^{LJ}(4,2)\nonumber
\end{align}
where for brevity the orbital indices have been suppressed, the spin indices $(I,J,L,M)$ are given in the Pauli basis, $v^{ML}$ is the generalized electron-electron interaction, and the polarizability is defined as,
\begin{align}\label{eq:bubble}
\chi^{IJ}_0(1,2)&=G(1,2)G(2,1^+)\\
&+G(1,3)G(4,1^+)\bar{\Gamma}(3,4;9,10)G(9,2)G(2^+,10).\nonumber
\end{align}
Here, $\bar{\Gamma}$ describes a multiple scattering process of two quasiparticles with the vertex, see Sec. II.B of Ref.~\onlinecite{lane2022identifying} for details. Assuming the material exhibits sufficient screening, the vertex correction $\bar{\Gamma}$ can be considered negligible and ignored. Then Eq.~\ref{eq:general_response} can be solved outright, producing a generalized RPA-type matrix equation
\begin{align}\label{eq:rpa}
\chi^{MN}(\mathbf{q},\omega)&= \left[ 1-\chi^{MI}_{0}(\mathbf{q},\omega)v^{IK}  \right]^{-1}  \chi^{KN}_{0}(\mathbf{q},\omega), \\
&= \left[ 1-\bar{F}^{MK}(\mathbf{q},\omega) \right]^{-1}  \chi^{KN}_{0}(\mathbf{q},\omega).\nonumber
\end{align}
We note that in order to solve for $\chi^{IJ}$ we have introduced the matrix inverse of $1-\bar{F}$. Therefore, extra care must be taken when interpreting the response function. For a system exhibiting an ordered phase, e.g. AFM order, the poles $1-\bar{F}$ generated in the various spin and orbital channels predict bosonic quasiparticles, such as magnons. In a non-ordered system, if $\chi^{IJ}(\mathbf{q},\omega=0)\gg 1$, then the ground state is unstable to a broken-symmetry phase. The specific charge and spin instabilities of the system can be made transparent by diagonalizing the kernel $\bar{F}$, 
\begin{align}\label{eq:eigenmodes}
\bar{F}=V(\mathbf{q},\omega) \Lambda_{F}(\mathbf{q},\omega)V^{-1}(\mathbf{q},\omega)
\end{align}
where $\Lambda_{F}$ is a diagonal matrix, and $V$ is the eigenvector. Then
\begin{align}
&\chi^{MN}(\mathbf{q},\omega)=\\
& V_{M\alpha}(\mathbf{q},\omega)\left[ 1-\Lambda^{\alpha}_{F}(\mathbf{q},\omega) \right]^{-1}V_{\alpha K}^{-1}(\mathbf{q},\omega)  \chi^{KN}_{0~}(\mathbf{q},\omega). \nonumber
\end{align}
where $\alpha$ enumerates the instability `bands'. Now, as the instability strength $\Lambda_{F}^{\alpha}(\mathbf{q},\omega=0)$ approaches $1$, $\chi^{IJ}(\mathbf{q},\omega=0)$ becomes singular, or physically, the ground state becomes unstable to an ordered phase. Additionally, the momentum satisfying $\Lambda^{\alpha_{max}}_{F}(\mathbf{Q},\omega=0)= 1$, where $\alpha_{max}$ is the index of the maximum instability band, is the propagating vector $\mathbf{Q}$ of the emerging Stoner instability in the multiorbital spin-dependent system. The character of this instability may then be obtained by analyzing the associated eigenvectors, $V$.

Figure~\ref{fig:flucutations} (a) presents the momentum dependence of the maximum instability $\Lambda^{0}_{F}(\mathbf{q},\omega=0)$ for pristine LaNiO$_2$ in the NM phase for various slices along $q_z$ in the Brillouin zone. The overall peak structure in $\Lambda^{0}_{F}$ follows the folded Fermi surface of the Ni-$3d_{x^2-y^2}$ band,  with the main ridges displaying minimal $q_z$ dispersion. The dominant peaks are concentrated around the M$-$A edge of the Brillouin zone, with weaker satellites along the path connecting the edge and zone center. The momenta  $\mathbf{Q}^*$ of the largest instability (blue and red `x' ) in each slice is found to evolve along the $q_z$-axis, taking positions at $(\pi-\delta,\pi,0)$, $(\pi-\delta, \pi,q_z)$, $(\pi-\delta, \pi-\eta,q_z)$, $(\pi-\xi, \pi-\xi,q_z)$, and $(\pi,\pi,\pi)$. Interestingly, these momentum points yield nearly degenerate instability strengths, where only $0.0178$ separates the momenta of maximum $\Lambda^{0}_{F}$ in the $q_z=0$ and $q_z=\pi/c$ planes. A similar near degeneracy is found for various in-plane momenta surrounding the M$-$A edge of the zone [Fig.~\ref{fig:flucutations} (b)], with an instability strength difference of $0.0889$ and $0.0023$ between $Q_1$ and $Q_2$ in the $\Gamma$ and Z planes, respectively. By analyzing the eigenvectors at the various marked $\mathbf{Q}^*$ points, we find magnetic fluctuations to dominate by an order of magnitude over the charge sector. Moreover, these magnetic fluctuations are singularly composed of intra-orbital Ni-$d_{x^2-y^2}$ character, with the charge channel supported by inter-orbital Ni-$t_{2g}/e_{g}$ hybridization. These results clearly imply that the leading G-type antiferromagnetic instability [$\mathbf{Q}^*=(\pi,\pi,\pi)$] is virtually degenerate with a dense manifold of 2D and 3D incommensurate magnetic stripe orders. This is consistent with total energy DFT and DFT+DMFT results yielding a myriad of nearly degenerate magnetic configurations that lower the total energy with respect to the NM phase\cite{Gu2020,lechermann2021doping,wan2021exchange,liu2020electronic,jung2020antiferromagnetic,zhang2021magnetic}.

In Fig.~\ref{fig:flucutations} (a) and (b) the momentum dependence of the maximum instability is found to be quite flat throughout the Brillouin zone, producing a clear pile-up of various magnetic configurations within an infinitesimally small instability strength. To make this statement more precise and accurately count the total number of competing magnetic configurations, we introduce the density of instabilities $\lambda(\Omega)$, where $\lambda(\Omega)\delta\Omega$ is the number of instabilities in the system whose strengths lie in the range from $\Omega$ to $\Omega+\delta\Omega$. That is, $\lambda(\Omega)$ is defined as 
\begin{align}
\lambda(\Omega)=\frac{1}{N_{\mathbf{q}}}\sum_{\alpha \mathbf{q}}\delta(\Lambda^{\alpha}_{F}(\mathbf{q},0)-\Omega),
\end{align}
where $\Omega$ is the instability strength and $\alpha$ enumerates the instability eigenvalues defined in Eq.~\ref{eq:eigenmodes}. We further emphasize that $\lambda(\Omega)$ contains the instability information for all eigenvalues, not just for the maximum.

Figure~\ref{fig:flucutations} (c) shows the density of instabilities for pristine LaNiO$_2$ in the NM phase, along with an inset illustrating the region of origin of the various key features. The spectrum reveals two clear Van Hove-like singularities, one close to the maximum instability ($\mathbb{I}$) and the other in the body just above $0.8$ instability strength ($\mathbb{II}$), along with a large step edge at the bottom of the spectrum ($\mathbb{III}$). As was pointed out by L\'{e}on Van Hove\cite{van1953occurrence} in 1953, the appearance of singular features in the density of states of either electrons or phonons is intimately connected to the topology of the underlying band structure. Here, the presence of these singularities implies the existence of saddle points in the momentum dependent instability $\Lambda_F^{0}$. For example, peak $\mathbb{I}$ originates from the very subtle change of $\Lambda^{0}_{F}$ from being a local minima at $(\pi,\pi,0)$ to a global maximum at $(\pi,\pi,\pi)$ [Fig.~\ref{fig:flucutations} (a)], implying the existence of a critical $q_{z}^*$ where the concavity changes sign (saddle point). Peak $\mathbb{II}$ emerges from the flat plateau between instability band edges, marked in the inset as the tips of the four pointed star. The relative placement of the singularities, along with the step edge $\mathbb{III}$, suggests that the fluctuations in LaNiO$_2$ are 3D in nature. However, since peak $\mathbb{I}$ is in very close proximity to the maximum instability, the system is very close to a 3D-2D transition.

Much like the presence of a Van Hove singularity near the Fermi level can modify and enhance correlated physics of an interacting electron liquid\cite{irkhin2001effects,schulz1987superconductivity,markiewicz1997survey,bok2012superconductivity,newns1992saddle,zheleznyak1997parquet,zanchi1998weakly}, a similar `Van Hove scenario' arises when a saddle point in the density of instabilities nearly fulfills the Stoner criteria, $\Lambda_F\sim 1$\cite{markiewicz2017entropic}. In the latter case, a large population of charge and magnetic instabilities with different propagating $\mathbf{q}$-vectors are able to interact and compete with one another, thus requiring the addition of vertex corrections at the two-particle level to accurately capture the low-temperature physics\cite{moriya2012spin,tremblay2012two}. When corrections are included, e.g. within the two-particle self-consistent approach or self-consistent renormalization theory, these mutual interactions between the various instabilities can suppress long-range order, leaving strong fluctuating, short-range correlations to dominate down to very low temperatures\cite{yushankhai2008self,markiewicz2017entropic,lonzarich1985effect,ortiz2022magnetic}, and generate a pseudogap in the single particle spectral function\cite{moukouri2000many,tremblay2006pseudogap}. This scenario is consistent with the lack of an observed Mott or AFM phase, and the presence of strong AFM fluctuations, local magnetism, and a pseudogap-like weakly insulating state in pristine LaNiO$_2$\cite{osada2021nickelate,zeng2022superconductivity,ortiz2022magnetic}. 

\begin{figure}[t!]
\includegraphics[width=0.99\columnwidth]{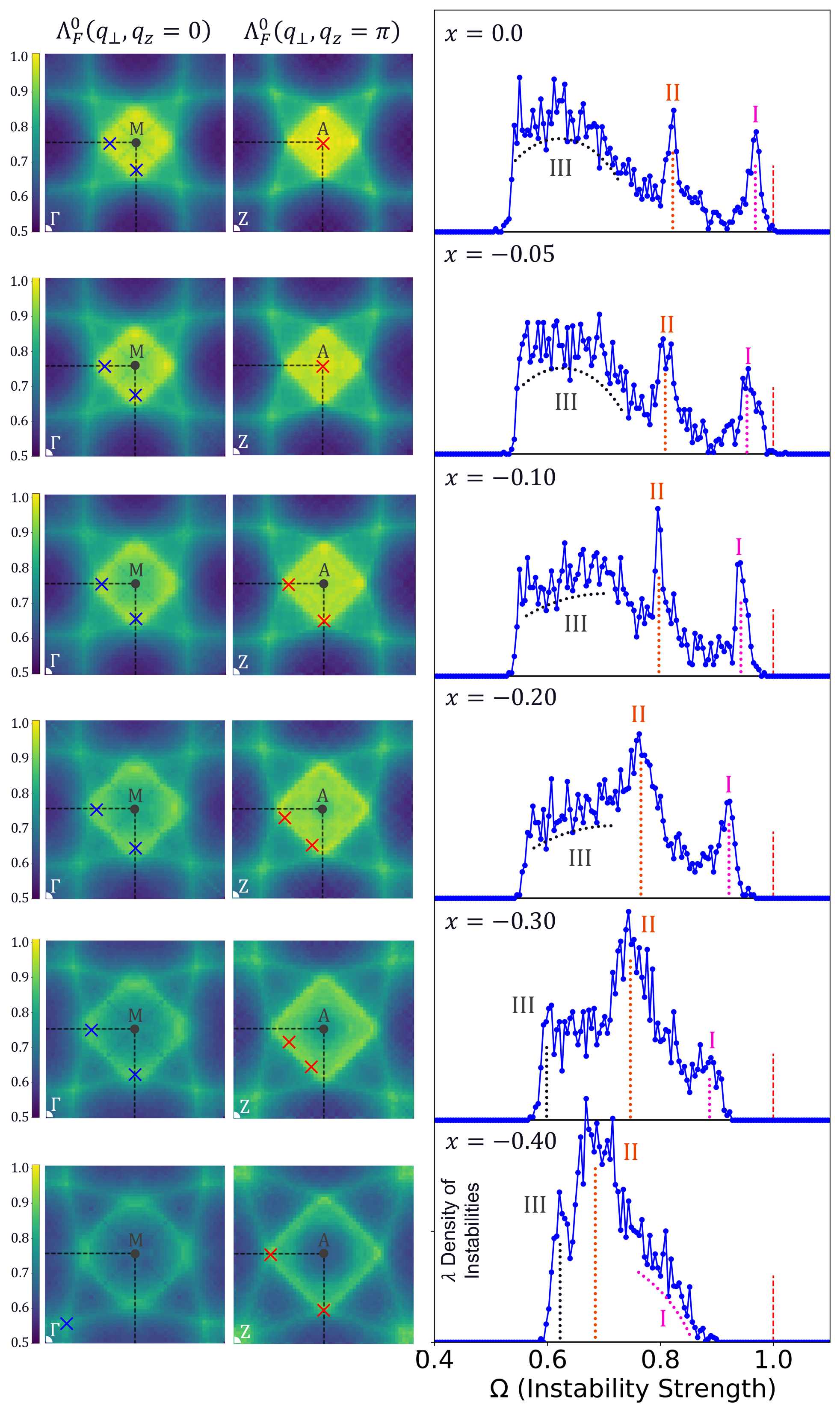}
\caption{(color online) Momentum dependence of the maximum instability $\Lambda^{0}_{F}(\mathbf{q},\omega=0)$ in the $\Gamma$ and Z planes of the Brillouin zone, along with the corresponding instability density of states of LaNiO$_2$ in the NM phase for various hole dopings. The various red and blue `x' mark the maximal and sub-maximal instability momenta $\mathbf{Q}^{*}$, respectively. The black dashed line denotes the boundary of the Brillouin zone. The two Van Hove singularities and step edge are indicated by Roman numerals $\mathbb{I}$, $\mathbb{II}$, and $\mathbb{III}$.} 
\label{fig:doping}
\end{figure}

\begin{figure*}[t!]
\includegraphics[width=0.99\textwidth]{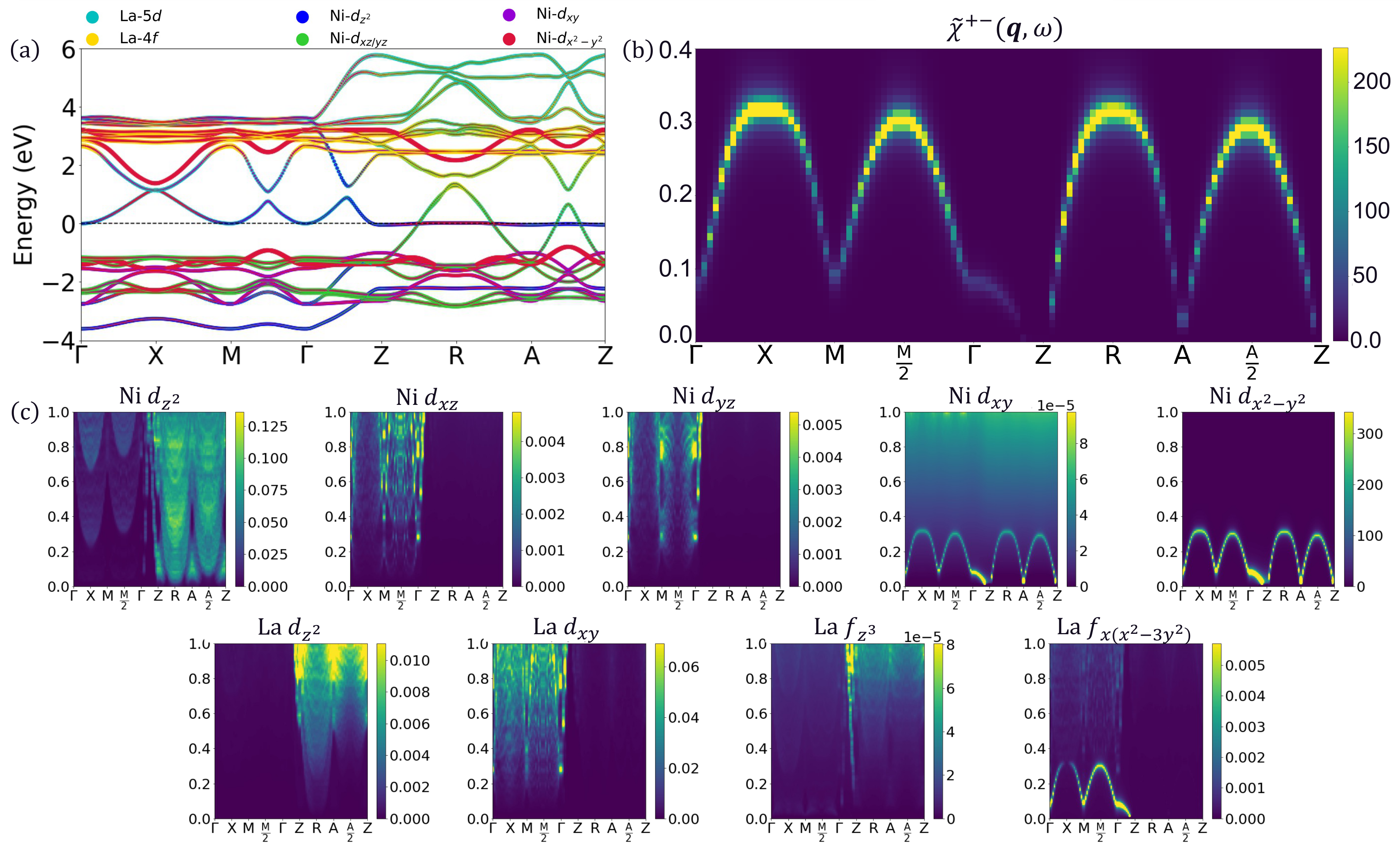}
\caption{(color online) (a) Electronic band dispersion for LaNiO$_2$ in the C-AFM phase. The size and color of the dots is proportional to the fractional weight of the various indicated site-resolved orbital projections. The observable (b) and orbitally decomposed (c) transverse spin susceptibility along the various high-symmetry lines in the NM Brillouin zone. } 
\label{fig:excitation}
\end{figure*}

Figure~\ref{fig:doping} shows the evolution of the momentum dependence of $\Lambda^{0}_{F}$ and the corresponding density of instabilities for LaNiO$_2$ under various hole dopings. As hole carriers are added to the system, the Ni-$3d_{x^2-y^2}$ Fermi surface sheet expands in volume, gradually reducing and eliminating	 the electron-like Fermi surface in the $k_z=\pi/c$ plane. Consequently, the areas of regions $\mathbb{I}$ and $\mathbb{II}$ in $\Lambda^{0}_{F}$, grow as shown in Fig.~\ref{fig:flucutations} (c), with increased hole concentration. Moreover, the concavity of $\Lambda^{0}_{F}$ at $(\pi,\pi,\pi)$ goes from negative to positive around 10\% hole doping, as illustrated by the momenta of maximal $\Lambda^{0}_{F}$ (red `x') changing locations from $(\pi,\pi,\pi)$ to $(\pi-\delta,\pi-\eta,\pi)$. This process is reflected in the density of instabilities where peak $\mathbb{I}$ transitions from a saddle point (x=0.0) to a step edge (x=0.30) Van Hove-like singularity, thereby precipitating an effective dimensionality reduction of the fluctuations from 3D to 2D. Curiously, this transition appears just before the sign change in the Hall coefficient at the start of the superconducting dome\cite{zeng2022superconductivity}.  Finally, at $x=0.4$, the leading edge of the density of instabilities softens, exhibiting a small number for instabilities suggesting a severe reduction in the competition between magnetic configurations.

The persistence of the peak $\mathbb{I}$ at or near the maximum instability edge for both the underdoped and overdoped regimes, suggests the preservation of strong competition between the magnetic states despite the systematic reduction in the fluctuation strength with doping. The 3D to quasi-2D transition in the nature of the fluctuations just before optimal doping makes LaNiO$_2$ distinct from the cuprates, which display predominantly 2D fluctuations for all hole dopings. Furthermore, this suggests that 2D magnetic fluctuations, in particular, are important for Cooper paring, with optimal $T_c$ arising from the delicate balance between the dimensionality and strength of the fluctuations. 

\section{Spin Excitation Spectra }
So far, we have examined the manifold of possible magnetic states that may be obtained in the pristine and doped LaNiO$_2$, along with an estimate of their competition. However, to gain insight into the magnetic excitation spectra of these phases, and thus try to connect the physical properties of LaNiO$_2$ with its electronic structure, we pick a representative realization for further scrutiny. Here, we choose the C-AFM state to analyze since it is a commensurate analog of the $(\pi,\pi,q_z^*)$ saddle point incommensurate order. To calculate the spin-flip excitation spectrum, we evaluate the imaginary part of the dynamical susceptibility [Eq.~\ref{eq:rpa}] for finite frequencies. The intensity spectra in the transverse (spin-flip) component $-\frac{1}{\pi}\text{Im}\tilde{\chi}^{+-}(\mathbf{q},\omega)$ of the effective observed susceptibility
\begin{align}
\tilde{\chi}^{+-}(\mathbf{q},\omega)=\sum_{\mu\nu}\chi^{+-}_{\mu\nu;\mu\nu}(\mathbf{q},\omega)
\end{align}
 are then compared to the RIXS results reported in Ref.~\onlinecite{lu2021magnetic}. By keeping the Ni-$d$, La-$d$, and La-$f$ orbitals, we decompose the spin-wave dispersion into the key atomic sites and orbitals contributing to the spin excitation.

Figure~\ref{fig:excitation} (a) presents the electronic band structure of LaNiO$_2$ in the C-AFM phase. The AFM state stabilizes in the Ni-$3d_{x^2-y^2}$ band with a gap of approximately 2 eV that opens up around the Fermi energy of the NM system. Notably, the partially filled Ni-$3d_{xy/yz}$ and Ni-$3d_{z^2}$ bands remain pinned to the Fermi level, where the $d_{z^2}$ derived band becomes essentially flat in the Z plane. Therefore, at this energy we can expect $\chi^{+-}$ to exhibit a mixture of local and itinerant magnetic excitations. 

Figure~\ref{fig:excitation} (b) shows $\text{Im}\tilde{\chi}^{+-}(\mathbf{q},\omega)$ along the various high-symmetry lines in the AFM Brillouin zone. Highly dispersive spin excitations are found following the spin-1/2 AFM magnons on a square lattice. Specifically, they disperse strongly with maxima at X and M/2, and linearly soften toward the AFM ordering wave vector at M in very good accord with the experimental spectra\cite{lu2021magnetic}. Our theoretical excitations display a gap in the $\Gamma$ plane compared to those in the plane through Z. This expected behavior is typical of infinite-layer perovskites, such as CaCuO$_2$\cite{peng2017influence}, where a pronounced in- and out-of-plane exchange coupling anisotropy can gap out magnetic excitations. By inspection of our electronic band structure, this gap can be attributed to the very small electron pockets at the Fermi level in the $\Gamma$-plane compared to those in the Z-plane. Additional small anisotropies within the $xy$-plane are also observed. Decomposing the spectrum into its orbital components, the Ni-$3d_{x^2-y^2}$ orbital is found to dominate, with minor contributions from Ni-$3d_{xy}$. Surprisingly, the La-$4f_{x(x^2-3y^2)}$ state has notable weight in the magnon spectrum in the $q_z=0$ plane, but none in the  $q_z=\pi/c$ plane. The partially filled orbitals produce a continuum of magnetic excitations in either the $\Gamma$ or the Z plane. Namely, the nearly flat Ni-$3d_{z^2}$ band at the Fermi level generates a strong, highly dispersive continuum of transverse spin excitations. The existence of orbital character beyond Ni-$3d_{x^2-y^2}$ suggests the importance of nickel-rare earth hybridization in shaping the magnetic excitations, and could be contributing to the long range behavior of the Heisenberg exchange parameters. This behavior is to be contrasted with the curpates, where spin excitations are highly localized on the copper atoms\cite{kastner1998magnetic}. 

\section{Discussion}
Our analysis of the charge and magnetic response of the nickelates supports the presence of strong experimentally observed magnetic fluctuations in this material \cite{hayward1999sodium,hayward2003synthesis,ZhaoPRL2021,Lu2021science,rossi2022broken,Krieger2021,Tam2021,2021arXiv211113668O,Leonov2020,Kitatani2020}. We find that the nickelates host a large number of nearly degenerate, homogeneous magnetic (e.g. C- and G-AFM) as well as incommensurate stripe configurations. The large number of competing spin and charge orders suggested by our modeling, could lead to an exotic form of intertwined order in LaNiO$_{2}$, which could play a significant role in driving superconductivity, as suggested for other unconventional superconductors~\cite{Kievelson2003,fradkin2015colloquium,dagotto2005complexity}. The strong fluctuations we identified in the pristine phase could explain why superconductivity is observed in undoped LaNiO$_2$~\cite{rossi2022broken,osada2021nickelate}.

The existence of dispersive spin-wave excitations in pristine LaNiO$_2$ in the absence of long range AFM order is a clear signature of local magnetism, which is seen commonly in various strongly correlated materials\cite{le2011intense,le2013dispersive,lu2022spin,horigane2016spin}. The width of our theoretical magnon spectrum in the ordered C-AFM system ($0.32$ eV) is comparable to the corresponding experimental RIXS value of $\sim 0.2$. The experimental dispersion also  displays a slight upturn in the acoustic mode as it approaches the zone center, compared to the theoretical spectrum\cite{lu2021magnetic}. Similar discrepancies between theory and experiment are also found in the doped cuprates\cite{coldea2001spin,le2011intense,le2013dispersive}. These results give further credence to our assertion that the pristine nickelates are analogous to the doped cuprates in that the cuprates support similar inhomogeneous magnetic stripe phases and dispersive spin-wave excitations in the ground state\cite{zhang2020competing,le2011intense,le2013dispersive}.

The coexistence of a Van Hove singularity in the electronic spectrum at the Fermi level and a similar singularity in the density of instabilities in the nickelates is intriguing. In the presence of a large number of active carriers along with an enormous number of competing magnetic instabilities, one could imagine the emergence of a complex fluctuating normal state through a coherent feedback between the two peaks associated with the aforementioned singularities in the electronic spectrum and the spectrum of instabilities. Notably, only a few reports of possible charge ordering in the nickelates have been posted~\cite{rossi2022broken,Krieger2021,Tam2021} and signatures of magnetic stripes have not been observed. In-plane strain might stabilize a particular magnetic stripe and yield local unidirectional spin-density waves, which would be readily accessible to local probes such as scanning-tunneling microscopy and nano-angle-resolved photoemission spectroscopy. 

\section{Concluding Remarks}\label{sec:conclusion}
Our study demonstrates that LaNiO$_2$ supports myriad competing, incommensurate spin fluctuations and magnetic excitations that are spread over multiple atomic sites. Interestingly, the reduction in the dimensionality of the magnetic fluctuations (3D to 2D) is found to coincide with the emergence of superconductivity. With further hole doping, fluctuations weaken, and superconductivity disappears, suggesting a tradeoff between dimensionality and strength of magnetic fluctuations in controlling the value of $T_c$. This behavior is similar to the doped cuprates where 2D inhomogeneous magnetic stripes are manifest already in the ground state\cite{zhang2020competing}. Our study gives insight into the nature of strong correlations in the nickelates and provides a pathway for investigating complex correlated materials more generally. 

\section*{Acknowledgements}  
\begin{acknowledgments}
The authors would like to thank Dr. Peter Mistark for many fruitful discussions during the early development of this work. The work at Los  Alamos National  Laboratory was carried out under the auspices of the US Department of Energy (DOE) National Nuclear Security Administration under Contract No. 89233218CNA000001. It was supported by the LANL LDRD Program, and in part by the Center for Integrated Nanotechnologies, a DOE BES user facility, in partnership with the LANL Institutional Computing Program for computational resources. Additional computations were performed at the National Energy Research Scientific Computing Center (NERSC), a U.S. Department of Energy Office of Science User Facility located at Lawrence Berkeley National Laboratory, operated under Contract No. DE-AC02-05CH11231 using NERSC award ERCAP0020494. The work at Tulane University was supported by the start-up funding from Tulane University, the Cypress Computational Cluster at Tulane, the Extreme Science and Engineering Discovery Environment (XSEDE), the DOE Energy Frontier Research Center (development and applications of density functional theory): Center for the Computational Design of Functional Layered Materials (DE-SC0012575), the DOE, Office of Science, Basic Energy Sciences Grant DE-SC0019350, and the National Energy Research Scientific Computing Center. The work at Northeastern University was supported by the US Department of Energy, Office of Science, Basic Energy Sciences Grant No. DE-SC0022216 and benefited from Northeastern University’s Advanced Scientific Computation Center and the Discovery Cluster and the National Energy Research Scientific Computing Center through DOE Grant No. DE-AC02-05CH11231. B.B. acknowledges support from the COST Action CA16218. 
\end{acknowledgments}

\appendix
\section{Scalable Spin and Orbital Resolved Lindhard Susceptibility Calculations}\label{Appendix:Lindhard}
To calculate the polarizability $\chi_0(\mathbf{q},\omega)$, we assume a non-interacting ground state, which allows us to replace the dressed single particle Green's function in Eq.~\ref{eq:bubble} with its non-interacting counterpart, 
\begin{align}
G_{0~\mu b , \nu^\prime a^{\prime}}(\mathbf{\kappa},i\omega_{n})=\sum_{i}\frac{V_{(\mu b ),i}V^{*}_{(\nu^\prime a^{\prime}),i}}{i\omega_{n}-\varepsilon_{i}},
\end{align}
where Greek and Latin letters denote the orbital and spin degrees of freedom, respectively, $i\omega_{n}$ is the Matsubara frequency and $V_{(\mu b ),i}=\braket{\mu b |i}$  are the matrix elements connecting the orbital-spin and the band spaces found by diagonalizing the Hamiltonian. By introducing the material specific details in this manner, our approach is able to utilize model or {\it ab initio} derived tight-binding Hamiltonians\cite{marzari2012maximally}. Using $G_{0}$, the polarization function [Eq.~\ref{eq:bubble}] can be simplified by performing the Matsubara frequency summation and  analytically continuing $i\omega_{n}\rightarrow \omega+i\delta $, for $\delta\rightarrow 0^{+}$, yielding a multiorbital, spin-dependent Lindhard susceptibility of the form   
\begin{widetext}
\begin{align}
\chi^{IJ}_{0~\nu^\prime \nu;\mu^\prime \mu}(\mathbf{q},\omega)&=
\sum_{\substack{ab\\ a^{\prime}b^{\prime}}} 
\sigma^{I}_{a^{\prime}b^{\prime}}
\sigma^{J}_{ab}
\frac{1}{N_\mathbf{k}}\sum_{\mathbf{k}}\sum_{ij}
 V^{\mathbf{k}+\mathbf{q}}_{(\nu^{\prime}b^{\prime})i}
 V^{\dagger~\mathbf{k}+\mathbf{q}}_{i(\mu a)}
 V^{\mathbf{k}}_{(\nu b)j}
 V^{\dagger~\mathbf{k}}_{j(\mu^{\prime}a^{\prime})} 
\frac{f(\varepsilon^{j}_{\mathbf{k}})-f(\varepsilon^{i}_{\mathbf{k}+\mathbf{q}})}{\omega+\varepsilon^{j}_{\mathbf{k}}-\varepsilon^{i}_{\mathbf{k}+\mathbf{q}}+i\delta}.
\end{align}
\end{widetext}
In evaluating this propagator computations scale linearly in $\mathbf{q}$, $\mathbf{k}$, and $\omega$ as $N_{\mathbf{q}} \times N_{\mathbf{k}} \times N_{\omega}$ making calculations over dense meshes out of reach. However, to obtain the magnetic excitation spectrum for a real materials with many bands, dense meshes are required to sufficiently capture the magnon dispersions on the small energy and momentum scales. Therefore, to enable calculations on real materials we can rewrite $\chi^{IJ}_{0}$ in the spectral representation by multiplying by a judicious choice of unity to reduce the computational complexity, similar to Ref.~\onlinecite{miyake2000efficient,shishkin2006implementation}. That is, by inserting an integral over the Dirac delta function,
\begin{widetext}
\begin{align}
\chi^{IJ}_{0~\nu^\prime \nu;\mu^\prime \mu}(\mathbf{q},\omega)&=
\sum_{\substack{\alpha\beta\\ \alpha^{\prime}\beta^{\prime}}} 
\sigma^{I}_{\alpha^{\prime}\beta^{\prime}}
\sigma^{J}_{\alpha\beta}
\frac{1}{N_\mathbf{k}}\sum_{\mathbf{k}}\sum_{ij}
 V^{\mathbf{k}+\mathbf{q}}_{(\nu^{\prime}\beta^{\prime})i}
 V^{\dagger~\mathbf{k}+\mathbf{q}}_{i(\mu\alpha)}
 V^{\mathbf{k}}_{(\nu\beta)j}
 V^{\dagger~\mathbf{k}}_{j(\mu^{\prime}\alpha^{\prime})} 
\int_{-\infty}^{+\infty}dX
\frac{f(\varepsilon^{j}_{\mathbf{k}})-f(\varepsilon^{i}_{\mathbf{k}+\mathbf{q}})}{\omega-X+i\delta}\delta(X-[\varepsilon^{i}_{\mathbf{k}+\mathbf{q}}-\varepsilon^{j}_{\mathbf{k}}])
\end{align}
\end{widetext}
we can reorder the integral and the sum over $\mathbf{k}$, yielding
\begin{widetext}
\begin{align}
\hat{\chi}^{IJ}_{0~\nu^\prime \nu;\mu^\prime \mu}(\mathbf{q},X)&=
\sum_{\substack{ab\\ a^{\prime} b^{\prime}}} 
\sigma^{I}_{a^{\prime}b^{\prime}}
\sigma^{J}_{ab}
\frac{1}{N_\mathbf{k}}\sum_{\mathbf{k}}\sum_{ij}
 V^{\mathbf{k}+\mathbf{q}}_{(\nu^{\prime}b^{\prime})i}
 V^{\dagger~\mathbf{k}+\mathbf{q}}_{i(\mu a)}
 V^{\mathbf{k}}_{(\nu b)j}
 V^{\dagger~\mathbf{k}}_{j(\mu^{\prime} a^{\prime})} 
(f(\varepsilon^{j}_{\mathbf{k}})-f(\varepsilon^{i}_{\mathbf{k}+\mathbf{q}}))\delta(X-[\varepsilon^{i}_{\mathbf{k}+\mathbf{q}}-\varepsilon^{j}_{\mathbf{k}}]),
\end{align}
\end{widetext}
and
\begin{align}
\chi^{IJ}_{0~\nu^\prime \nu;\mu^\prime \mu}(\mathbf{q},\omega)&=
\int_{-\infty}^{+\infty}
\frac{ \hat{\chi}^{IJ}_{0~\nu^\prime \nu;\mu^\prime \mu}(\mathbf{q},X)}{\omega-X+i\delta}dX.
\end{align}
Lastly, when discretizing and implementing the above expressions, the limit of bins representation of the Dirac delta function was employed,
\begin{align}
\delta(z)=\lim_{dX \to 0^+} \frac{\theta(z+\frac{dX}{2})-\theta(z-\frac{dX}{2}) }{dX},
\end{align}
where $dX$ is the bin width of the auxiliary $X$-mesh, and since $\hat{\chi}^{IJ}_{0}$ has compact support for a finite set of bands, we may take the bounds of the integral to be $X_{max}$ and $X_{min}$ such that $\hat{\chi}^{IJ}_{0}(\mathbf{q},\omega)$ is zero for all $X>X_{max} (X<X_{min})$.

Now, the calculation of $\chi^{IJ}_{0}$ is split into two parts: the first, a binning process over the excitation energies $\varepsilon^{i}_{\mathbf{k}+\mathbf{q}}-\varepsilon^{j}_{\mathbf{k}}$ for a given $\mathbf{q}$, and second, a Hilbert transform of $\hat{\chi}^{IJ}_{0}$ to recover $\chi^{IJ}_{0}$. This process of splitting the calculation of the density-density propagator has overall reduced the computational complexity from $N_{\mathbf{q}} \times N_{\mathbf{k}} \times N_{\omega}$ to $N_\mathbf{q} (N_\omega N_X + N_\mathbf{k})$ yielding a significant computational advantage when $N_X <N_{\mathbf{k}}$, which is usually true. Therefore, this scheme enables us to push our calculations to dense $\mathbf{q}(\mathbf{k})$ and $\omega$ meshes.

\bibliography{nickelates_Refs}

\end{document}